\documentstyle[aps,multicol]{revtex}
\input epsf
\begin{document}
\draft
\widetext
\title{Non-Langevin behaviour of the uncompensated magnetization in nanoparticles of
artificial ferritin.} 
\author{C. Gilles, P. Bonville}
\address{CEA, C.E. Saclay, Service de Physique de l'Etat Condens\'e, 91191 Gif
-sur-Yvette, France}
\author{K. K. W. Wong, S. Mann}
\address{School of Chemistry, University of Bristol, England}
\date{\today}\maketitle\widetext
\begin{abstract}
\leftskip 54.8pt
\rightskip 54.8pt
The magnetic behaviour of nanoparticles of antiferromagnetic artificial ferritin
has been investigated by $^{57}$Fe M\"ossbauer absorption spectroscopy and magnetization 
measurements, in the temperature range 2.5\,K-250\,K and with magnetic fields up to 7\,T.
Samples containing nanoparticles with an average number of $^{57}$Fe atoms ranging from 
400 to 2500 were studied. By analysing the magnetic susceptibility and zero field
M\"ossbauer data, the anisotropy energy per unit volume is found, in agreement with some of
the earlier studies, to have a value typical for ferric oxides, i.e. a few 10$^5$\,ergs/cm$^3$.
By comparing the results of the two experimental methods at higher fields, we 
show that, contrary to what is currently assumed, the uncompensated magnetisation of the
ferritin cores in the superparamagnetic regime does not follow a Langevin law. For magnetic 
fields below the spin-flop field, we propose an approximate law for the field and 
temperature variation of the uncompensated magnetization, which was early evoked by N\'eel but 
has so far never been applied to real antiferromagnetic systems. 
More generally, this approach should apply to randomly oriented antiferromagnetic 
nanoparticles systems with weak uncompensated moments.\par
PACS: 75.50 Tt; 75.50 Ee; 76.80 +y
\end{abstract} 

\begin{multicols}{2}
\section{Introduction}\label{intro}
Natural ferritin is the iron-storage protein of animals, plants and bacteria. 
It is composed of a ferrihydrite-like core with formula (FeOOH)$_8$(FeOH$_2$PO$_4$)
about 7\,nm in diameter and containing up to about 4500 Fe atoms \cite{stpier},
surrounded by a 12\,nm diameter multisubunit protein shell.
By a suitable chemical synthesis process, it is possible to reconstitute the 
ferrihydrite ferritin
core inside the empty protein shell (in this case obtained from horse spleen ferritin
by reductive dissolution of the native core) \cite{mann}.
This way, one can monitor the amount of iron available for the build up of the core, 
and a better control of the average core size is obtained.
A number of investigations have been performed, both on natural and artificial ferritin,
using either $^{57}$Fe M\"ossbauer absorption spectroscopy 
\cite{bell,bauminger,pier,pankhurst,hunt} or magnetization measurements 
\cite{kilcoyne,berk,brooks,harris}. They have shown that the magnetic structure of the 
Fe$^{3+}$ ions in the core is probably antiferromagnetic, and that the particles possess
a (small) uncompensated magnetic moment. However, the interpretation of the data obtained
by the two techniques is somewhat contradictory. Indeed, on one side, the M\"ossbauer 
studies with 
applied magnetic field \cite{pier,pankhurst,hunt} reveal that, for randomly oriented
ferritin samples, the Fe$^{3+}$ magnetic moments in the cores remain close to their easy
axes, even in a strong field of 14\,T. On the other side, the magnetic measurements
have been interpreted so far by assuming that the uncompensated magnetization follows a 
Langevin law, which implicitly assumes the Fe$^{3+}$ moments are free to rotate.
Besides its fundamental interest, the
understanding of the magnetic properties of ferritin is of importance for NMR relaxometry
\cite{brooks} and biophysical purposes as it is a good contrasting agent for NMR imaging.
The ferritin core has also recently arisen great interest because it has been thought to be 
a good candidate for the observation of quantum tunneling of the magnetization
\cite{barbara}, and various experimental studies, at low temperature, have claimed 
to have observed this tunneling \cite{awschalom,gider,tejada}.\par

The aim of the present work is to reexamine the problem of the uncompensated 
magnetization in
ferritin, by comparing $^{57}$Fe M\"ossbauer data, both in zero field and in applied
fields up to 7\,T, with magnetization data obtained for the same samples.
We show that, by taking proper account of the crystalline anisotropy in an antiferromagnetic
structure, one obtains a coherent interpretation of the results of the two techniques.
We present a new interpretation of the superparamagnetic behaviour of an ensemble of 
randomly oriented antiferromagnetic particles, in terms of a non-Langevin 
law for the uncompensated magnetisation, for
applied magnetic fields lower than the spin-flop field. This approach, which was evoked
by N\'eel in an early work \cite{neel0}, has however never been considered in the modern and
quantitative magnetic investigations of ferritin. Another motivation of our work is
to examine the influence of the mean iron content of the (artificial) ferritin core (ranging 
from 400 to 2500 Fe atoms) on the physical properties, such as the anisotropy energy, 
or the N\'eel temperature.\par
The paper is organized as follows: section \ref{echant} contains the description of the 
artificial ferritin samples, section \ref{superpara} briefly recalls the superparamagnetic
behaviour of antiferromagnetic particles, both from the point of view of magnetometry and
$^{57}$Fe M\"ossbauer spectroscopy, sections \ref{zfc} and \ref{moss} contain respectively the
FC-ZFC magnetic susceptibility and zero field M\"ossbauer measurements, and the high field 
M\"ossbauer
data; in section \ref{exchange}, the influence of anisotropy in uncompensated antiferromagnets
is examined, and section \ref{aimantation} reports the magnetisation experiments and the new
interpretation we propose.

\section{Sample characterization and experimental techniques}\label{echant}
In order to investigate the effect of decreasing both the particle size
and the mean number of Fe atoms per ferritin core, the experiments were performed 
with a set of artificial ferritin samples with different Fe loadings. 
Five samples were studied, having an average iron loading ranging from 400 to 2500 Fe 
ions per ferritin core.
In order to obtain a good signal to noise ratio in the M\"ossbauer spectra, 
the Mohr salt used as the starting material was prepared with iron 95\%
enriched in $^{57}$Fe. 
The protein concentration was determined by the Lowry method and the mean iron content
by atomic adsorption analysis. 
In the following, the samples will be labelled according to the mean number of Fe atoms
per core determined by this latter method.
The samples consist of a solution of artificial ferritin with a concentration
of about 10\,mg/ml, that is ten times smaller than the concentration of commercial
ferritin (about 100\,mg/ml).
Then, the distance between particles in the solution is about 50\,nm, which corresponds 
to a dipolar field of magnitude 10\,mG (at saturation of the uncompensated moments). 
Our samples can therefore be considered as an ensemble of non-interacting particles.
Pictures taken by transmission electron microscopy (TEM) show the cores to be discrete
and roughly spherical in shape. Size histograms
have been established by measuring the diameter of 500 particles taken from
different parts of the grid. For each sample, the diameter distribution is rather 
narrow and can be fitted to a lognormal 
shape, with a mean diameter d$_0$ ranging from 4\,nm for the particles with 400 Fe atoms per
core to 5.7\,nm for the particles with 2571 Fe atoms per core, and a standard 
diameter deviation $\sigma$ about 0.15. Electron diffraction patterns, performed on
the particles with 2571 Fe atoms per core, give d-spacings corresponding to well-ordered 
ferrihydrite.

The magnetisation measurements were made with a commercial SQUID magnetometer
in magnetic fields up to 5.5\,T in the temperature range 2.5\,K-250\,K. 
For all temperatures and fields, we have measured both the signal of the solution 
containing ferritin and the signal of the solution containing 
apoferritin (the empty protein shells) with the same concentration. After 
subtraction of the second signal from the first, we thus obtain the signal due 
only to the ferritin cores. For the low field (80\,G) magnetisation measurements,
both Field Cooled (FC) and Zero Field Cooled (ZFC) branches were measured. 
In the Zero Field Cooled (ZFC) procedure, the sample was cooled in zero field from room 
temperature down to 2.5\,K, and for the Field Cooled (FC) branch, the 
sample was cooled with a field of 80\,G from room temperature to 2.5\,K; in both cases,
measurements proceeded on heating.\par

The $^{57}$Fe M\"ossbauer absorption spectroscopy experiments were performed both in zero
magnetic field and in magnetic fields up to 7\,T applied perpendicular to the
$\gamma$-ray propagation direction, between 4.2\,K and 90\,K.
The M\"ossbauer spectra were obtained using a $^{57}$Co:Rh source, mounted on
an electromagnetic drive with a triangular velocity signal.
The ferritin solution was placed in a copper holder covered with pure thin
Aluminium sheets, ensuring a good thermalization. 
For the in-field measurements, the ferritin solution is first frozen in zero field and 
therefore there is no preferential orientation of the magnetization of the particles.

\section{Superparamagnetism of antiferromagnetic particles}\label{superpara}
The Fe$^{3+}$ ion has a saturated magnetic moment $m_0$ =5\,$\mu_B$.
The magnetic structure of the ferritin cores is expected to be
antiferromagnetic, with a N\'eel temperature $T_N$ of the order of a few
hundred Kelvins. In zero external field, the preferential orientation of the 
two antiferromagnetic sublattices
in the core is determined by the crystalline anisotropy (``antiferromagnetic axis'').
For antiferromagnetic nanometric particles, 
one expects a small uncompensated magnetization which may arise from 
the core (presence of defects) and/or from the unpaired surface moments of the
particle \cite{neel1}. Due to the strong exchange interaction, the uncompensated
moments are aligned with the antiferromagnetic sublattice moments, except probably
at the surface due to broken exchange bonds \cite{kodama}.
At a given temperature $T$ below $T_{\rm N}$, the moments of the two sublattices fluctuate 
by crossing 
the anisotropy energy barrier: this is the superparamagnetic relaxation.
In the case of an axial anisotropy, the relaxation time for the reversal of the
direction of the magnetization of a particle with volume $V$ is described by an 
Arrhenius type equation first proposed by N\'eel \cite{neel2}:   
\begin{eqnarray}
\tau=\tau_0\ \exp({{KV}\over {k_BT}}),
\label{tauN}
\end{eqnarray}
where $K$ is the magnetic anisotropy energy per unit volume,
$\tau_0$ a microscopic relaxation time usually considered to be a constant with magnitude 
10$^{-10}$-10$^{-11}$s but varying in fact with $V$ and $T$ \cite{brown}.
Due to the particle size distribution present in
any real sample, there is a distribution of anisotropy barriers $KV$ which, according to
Eqn.\ref{tauN}, results in a broad distribution of relaxation times.
For a given measurement technique, with a characteristic time $\tau_m$, the blocking volume:
\begin{eqnarray}
V_b={{k_BT}\over K}\ln(\tau_m/\tau_0),
\label{Vb}
\end{eqnarray} 
defines two populations of particles: those with $V>V_b$ and those with $V<V_b$, whose
magnetization fluctuation time is respectively larger and smaller than $\tau_m$.
Using Eqn.(\ref{tauN}), one can also define a mean blocking temperature:
\begin{eqnarray}
T_b={{K \langle V \rangle }\over {k_B\ln(\tau_m/\tau_0)}},
\label{Tb}
\end{eqnarray} 
such that, for $T \ll T_b$ all the particles are in the ``frozen regime'' and for 
$T \gg T_b$ all the particles are in the ``superparamagnetic regime''.\par

For magnetization experiments, the characteristic time can be considered to be
the measurement time $\tau_{\chi}$, which is about 100\,s. 
In low fields, and for the superparamagnetic regime, N\'eel has shown that the 
magnetization is the sum of two contributions \cite{neel3}:
\begin{eqnarray}
M(H,T)=\chi_{\rm AF}H + M_{nc}(H,T),
\label{mh}
\end{eqnarray}
where $\chi_{\rm AF}$ is the susceptibility arising from the canting of the two 
antiferromagnetic sublattices, and $M_{nc}$ is the magnetization due to the 
uncompensated moments. At T=0 and in the case the magnetic field is perpendicular to the
antiferromagnetic axis, $\chi_{\rm AF} = \chi_\perp$ which, in the standard molecular 
field theory, is given by: $\chi_\perp = M_0/H_{\rm E}$, $M_0$ being the magnetization
of a sublattice and $H_{\rm E}$ the exchange field.
In the low field limit, the uncompensated magnetization follows a Curie 
law, irrespective of the strengh of the anisotropy \cite{neel3}.
At higher fields, the choice of an expression for $M_{nc}(H,T)$ is not a trivial problem,
and an important issue of the present work is the determination of the best approximation
for the field and temperature dependence of the uncompensated magnetization in 
antiferromagnetic particles.\par
The characteristic M\"ossbauer time is the Larmor period $\tau_{\rm L}$ associated with the
magnetic hyperfine interaction. 
For $^{57}$Fe, it is of the order of 5$\times$10$^{-9}$\,s. 
In the case of magnetic ordering of the Fe$^{3+}$ ions, there is a magnetic
hyperfine field ${\bf {H}_{hf}}$ at the nucleus, proportional to 
the Fe$^{3+}$ moment and directed opposite to it.
If the fluctuation time $\tau$ of the magnetization (and hence of the hyperfine field)
is longer than $\tau_{\rm L}$, the M\"ossbauer spectrum is a six-
line hyperfine field pattern; if it is smaller than $\tau_{\rm L}$, the M\"ossbauer spectrum
is a two-line quadrupolar pattern. In real systems, due to the broad distribution 
of $\tau$ values, a
zero field pattern consists of a superposition of magnetic and quadrupolar subspectra,
in a sizeable temperature range.
When a magnetic field is applied, each nucleus experiences an effective field 
${\bf H_{eff}}$, which is the vectorial sum of the applied field $\bf H$ and of the 
hyperfine field ${\bf H_{hf}}$:
\begin{eqnarray}
{\bf H_{eff}}={\bf H}+{\bf H_{hf}}.
\label{Heff}
\end{eqnarray}
If $\phi$ is the angle between the $\gamma$-ray propagation direction and the effective 
field direction, the intensity ratios of the outer to middle to inner pairs of lines of the
M\"ossbauer sextet are given by:
\begin{eqnarray}
3(1+\cos^2\phi):4\sin^2\phi:(1+\cos^2\phi).
\label{int}
\end{eqnarray}
If the effective field is aligned perpendicular to the $\gamma$-ray direction, the intensity
ratios are 3:4:1, and for a random orientation of the effective field, the intensity ratios 
are 3:2:1.

\section{Low field susceptibility and zero field M\"ossbauer experiments}\label{zfc}

\begin{figure}
\epsfxsize=7 cm
\centerline{\epsfbox{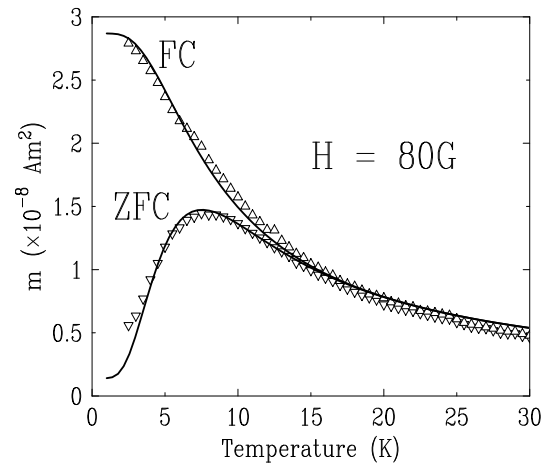}}
\vspace{0.5 cm}
\caption{\small\sl Thermal variation of the FC and ZFC magnetic susceptibility in the
artificial ferritin sample with a mean Fe loading of 410 atomes per core, measured in
a field of 80\,G. The solid lines are fits using to Eqn.(\protect\ref{mzfc}) and a 
Gaussian distribution of $\mu_{nc}$ values.}
\label{fc410}
\end{figure}

The two FC and ZFC branches of the magnetic susceptibility for the particles with 410 Fe
atoms per core, measured in a field $H$=80\,G, are represented in Fig.\ref{fc410}.
The signal is almost entirely
due to the uncompensated moments, the antiferromagnetic susceptibility being
very weak as will be shown in section \ref{aimantation}. 

\noindent The shape of the curves is typical of an ensemble of non-interacting 
relaxing moments with
a distribution of anisotropy barriers \cite{gittleman}. The peak temperature of the ZFC curve
for the particles
with 410 Fe atoms per core is $T_{\rm peak} \simeq$7.5\,K, and the irreversibility 
temperature 
at which the FC and ZFC branches join is $T_{\rm irr} \simeq$ 15\,K.
To interpret these curves, 
we follow here the model of Ref.\onlinecite{gittleman}. Assuming that, for a given particle
volume, the uncompensated moment $\mu_{nc}$ is a known function of $V$, the thermal 
variation of the FC and ZFC magnetic moment writes:
\begin{eqnarray} 
m_{nc}&(&T)={H\over 3k_BT} \int_{V_{min}}^{V_b(T)} \mu_{nc}(V)^2 f(V) dV + \cr
&\eta& {H\over 3K} \int_{V_b(T)}^{V_{max}} \mu_{nc}(V)^2 {f(V) \over V} dV,
\label{mzfc}
\end{eqnarray} 
where $f(V)$ is the log-normal volume distribution function, $V_b(T)$ the blocking volume 
associated with $\tau_{\chi}$, and with $\tau_0=10^{-11}$\,s, and $\eta$=0 for the ZFC branch and 
$\ln{{\tau_{\chi}} \over {\tau_0}}$ for the FC branch. However, in antiferromagnetic
particles, the relationship between volume and uncompensated moment is not well known.
The most probable is that, for a given volume, there is a distribution of uncompensated
moment values. According to an argument due to N\'eel \cite{neel1}, the mean value of this
distribution is proportional to $N^p$, where $N$ is the number of Fe atoms in the particle
and p=1/2, 1/3 or 2/3 in the case respectively of a random distribution of the uncompensated 
moments in the volume, at the surface, or 
in the presence of regular ``active'' (ferromagnetic) planes at the surface. It appears that 
the fit of the experimental FC and ZFC curves using Eqn.\ref{mzfc} is not very sensitive 
to the particular shape
of the distribution of uncompensated moments, nor to the choice of a particular $p$ value.
The solid lines in Fig.\ref{fc410}, which reproduce quite well
the experimental data, were obtained by introducing a Gaussian distribution of $\mu_{nc}$ 
values into expression (\ref{mzfc}). 

\begin{figure}
\epsfxsize=7 cm
\centerline{\epsfbox{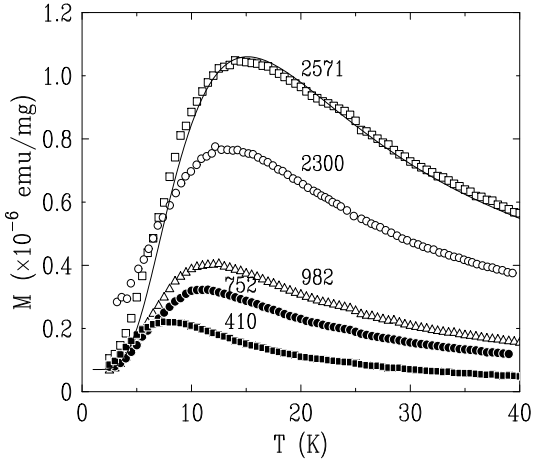}}
\vspace{0.5 cm}
\caption{\small\sl Thermal variation of the magnetization of artificial ferritin
samples with different Fe loadings (the figure near each curve represents the
mean number of Fe atoms per particle) upon warming in a field of 50\,G after cooling
in zero field (ZFC curves).
The fit using Eqn.(\protect\ref{mzfc}) (solid lines) is shown only for the sample
with a loading of 2571 Fe atoms per core. The magnetization is expressed in emu per 
mg of protein.} 
\label{courbezfc}
\end{figure}

\noindent The mean value of this distribution is taken as:
$\langle \mu_{nc} \rangle = m_0 u \sqrt{N}$ (assuming the Fe atomic density $n=N/V$ is 
constant for a given particle set), and its standard deviation as
$\sigma_{nc}$=0.4 $\langle \mu_{nc} \rangle$. The 
distribution is restricted to positive $\mu_{nc}$ values. The fitted value of the
proportionality parameter $u$ is in the range 0.8--1 for the different particle sets, and the 
standard deviation of the diameter distribution was kept at the value $\sigma$=0.15 
obtained from the size histograms. This choice
of a $N^{1 \over 2}$ law for the mean $\mu_{nc}$ value corresponds to a volumic random 
distribution 
of the Fe atoms bearing the uncompensated moments but, as emphasized before, the data do not 
allow us to exclude the presence of uncompensated moments at the surface.
Nevertheless, these fits allow the anisotropy constant $K$ to be determined: we find that 
$K$ lies in the range 3--6 10$^5$\,ergs/cm$^3$ for the different particle sets, with a tendency
to increase as the particle size decreases. \par
\begin{figure}
\epsfxsize=7 cm
\centerline{\epsfbox{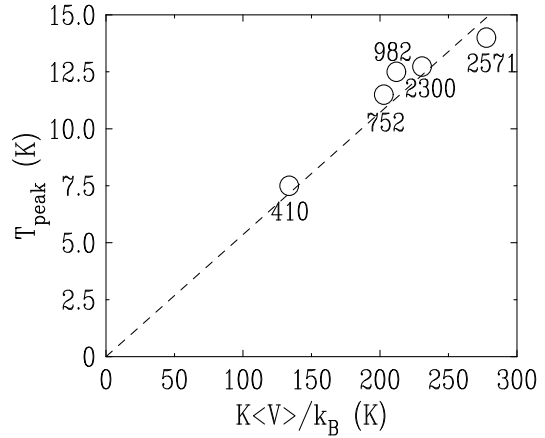}}
\vspace{0.5 cm}
\caption{\small\sl Dependence of T$_{\rm peak}$ of the ZFC curves on the mean 
anisotropy energy in artificial ferritin samples.} 
\label{varTpic}
\end{figure}

In Fig.\ref{courbezfc} is represented, for each sample, the behaviour of the ZFC
magnetization as a function of temperature (the fit is shown as a solid line only for the
particles with 2571 Fe atoms per core). The peak temperature increases with the particle 
size, and it varies approximately linearly with the mean barrier energy $K \langle V \rangle$,
as can 
be seen in Fig.\ref{varTpic}. As, in antiferromagnetic particles, the shape anisotropy
is very small \cite{bocq}, this linear correlation is expected if the volumic anisotropy 
dominates over the surface anisotropy. Indeed, in this case,
$T_{\rm peak}$ can be shown to be proportionnal to the blocking temperature $T_b$ 
\cite{gittleman,sappey}:
\begin{eqnarray} 
T_{\rm peak}=g(\sigma)T_b,
\label{tpic}
\end{eqnarray} 
where $g(\sigma)$ is an increasing function of the standard diameter deviation $\sigma$
with $g(\sigma\rightarrow0)=1$. As $\sigma$ is relatively constant for the different particle
sets, the peak temperature must be proportional to the mean anisotropy barrier owing to
Eqn.\ref{Tb}.\par

For all samples, the FC curve joins the ZFC curve above an irreversibility temperature which
is about twice the peak temperature; therefore, for all particle sizes, the 
superparamagnetic regime associated with the magnetization measurement time
occurs above 30\,K.\par

\begin{figure}
\epsfxsize=7 cm
\centerline{\epsfbox{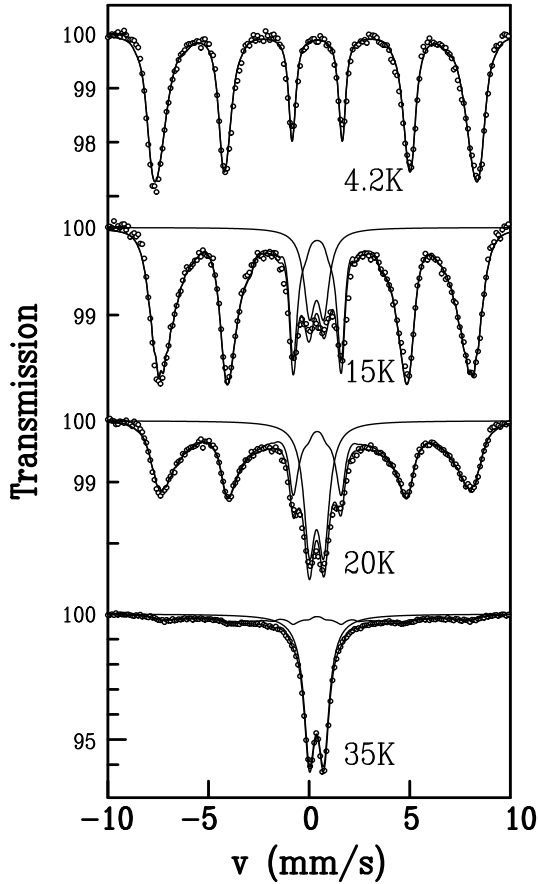}}
\vspace{0.5 cm}
\caption{\small\sl $^{57}$Fe M\"ossbauer absorption spectra at selected temperatures
in zero field in the
artificial ferritin sample with a mean Fe loading of 410 atoms per core.}
\label{mosschmpnul}
\end{figure}

Representative $^{57}$Fe M\"ossbauer absorption spectra in zero field for the
particles with 410 Fe atoms per core are shown in Fig.\ref{mosschmpnul}.
At 4.2\,K, a static magnetic hyperfine pattern is observed with a hyperfine
field $H_{hf} \simeq$ 490\,kOe, identical for all particle sets.
As temperature increases, the quadrupolar doublet progressively replaces the 
six-line magnetic pattern as the 
superparamagnetic fluctuation time of more and more particles becomes smaller than
$\tau_{\rm L}$, i.e. as the blocking volume given by Eqn.(\ref{Vb}) associated with 
the characteristic time
$\tau_{\rm L}$ becomes smaller. The spectra were fitted to a superposition of a 
quadrupolar hyperfine spectrum and of magnetic hyperfine spectra with a distribution
of hyperfine fields (histogram). 

\begin{figure}
\epsfxsize=7 cm
\centerline{\epsfbox{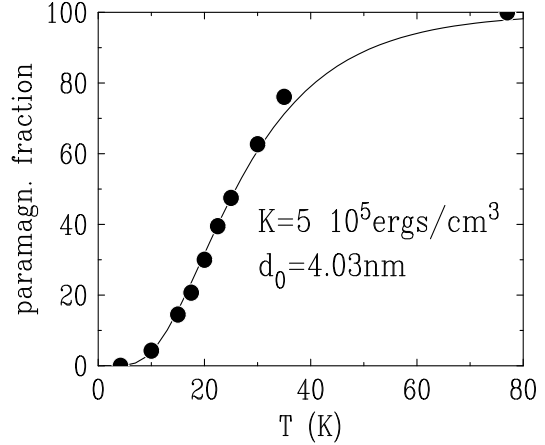}}
\vspace{0.5 cm}
\caption{\small\sl Thermal variation of the relative intensity of the quadrupolar
doublet in the artificial ferritin sample with a mean Fe loading of 410 atoms per core. 
The solid line is a fit using Eqn.(\protect\ref{parfract}).
The blocking temperature is 22\,K.}
\label{pourc}
\end{figure}

\noindent The thermal variation of the relative intensity $f_p(T)$
of the quadrupolar doublet derived from these fits is shown in Fig.\ref{pourc}.
This ``paramagnetic fraction'' can be
calculated in the frame of N\'eel's model of thermally activated fluctuations:
\begin{eqnarray}
f_p(T)={1 \over {\langle V \rangle}} \int_{V_{min}}^{V_b(T)} Vf(V)dV,
\label{parfract}
\end{eqnarray}
where $V_b(T)={{k_BT}\over K}\ln{\tau_{\rm L}\over \tau_0}$ and $\tau_0=10^{-11}$\,s.
The paramagnetic fraction is
independent of the values (or distribution) of the uncompensated moments as long as
one neglects the possible (but unknown in antiferromagnets) dependence of $\tau_0$ on 
$\mu_{nc}$. For all particle sets, the thermal variation of $f_p$ is well reproduced 
using Eqn.\ref{parfract} (solid line in Fig.\ref{pourc}),
with an anisotropy constant close to that derived from the fit of the FC-ZFC 
susceptibility curves.
The blocking temperature, defined as the temperature for which half of the particles
are in the superparamagnetic regime, increases with the particle mean size.
For all iron loadings, all the particles are in the superparamagnetic regime 
above 80-100\,K.

\section{High-field M\"ossbauer experiments}\label{moss}

According to the zero field M\"ossbauer data, for all particle sizes, the fluctuations of
the magnetization at 4.2\,K are slow with respect to the hyperfine Larmor frequency, and 
they are fast at 90\,K. In-field M\"ossbauer spectra at these temperatures have been 
recorded in natural ferritin and haemosiderin in Ref.\onlinecite{pier}. We repeated these
measurements in our artificial ferritin samples to try and detect the influence
of the particle size on the orientation of the magnetic moments in the frozen and 
paramagnetic regimes when a magnetic field up to 7\,T is applied. We actually found
that there is no drastic qualitative change of the in-field spectra for the different 
particle sets, and our data are similar to those recorded in natural ferritin.
We therefore only present the spectra for the particles with 410 Fe atoms per core, 
and our interpretation follows the same lines as that made in Ref.\onlinecite{pier}.

\subsection{Spectra in the frozen regime (T=4.2\,K)}
\begin{figure}
\epsfxsize=7 cm
\centerline{\epsfbox{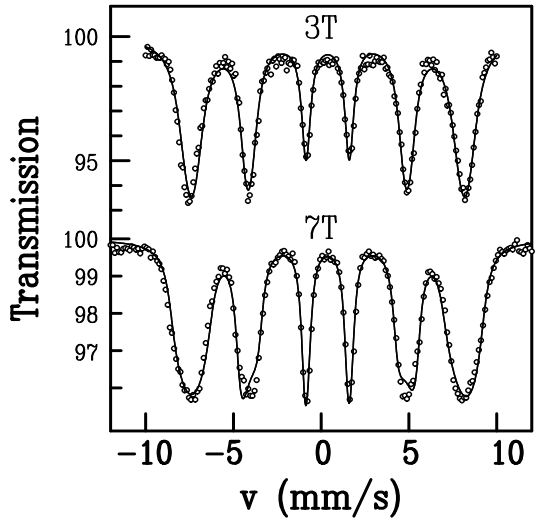}}
\vspace{0.5 cm}
\caption{\small\sl $^{57}$Fe M\"ossbauer absorption spectra in the frozen regime (4.2\,K)
in the artificial ferritin sample with a mean Fe loading of 410 atoms per core, 
with magnetic fields of
3\,T and 7\,T applied perpendicular to the $\gamma$-rays propagation direction.
The solid lines are a fit to a model with random orientation of the hyperfine fields.}
\label{moss4.2}
\end{figure}

The spectra at 4.2\,K for fields of 3\,T and 7\,T in these particles 
are shown in Fig.\ref{moss4.2}. They are resolved six line hyperfine spectra,
analogous to the zero-field spectrum (see top of Fig.\ref{mosschmpnul}).
The intensity ratios of the pairs of lines remain close to 3:2:1, even at 7\,T, which 
means that the field does not drastically modify the orientations of the individual Fe$^{3+}$
moments with respect to those in the randomly oriented zero-field configuration.
The spectra were then fitted assuming random orientations of the static hyperfine field 
${\bf H_{hf}}$, with fixed magnitude.
For a given angle $\theta_b$ between ${\bf H_{hf}}$ and the applied field,
the magnitude of the effective field ${\bf H_{eff}}={\bf H}+{\bf H_{hf}}$ experienced
by the nucleus is given by:
\begin{eqnarray}
H_{eff}=\{(H_{hf}\cos\theta_b+H)^2+H_{hf}^2\sin^2\theta_b\}^{1 \over 2}.
\label{Hefftetab}
\end{eqnarray}
The orientational disorder of $\bf H_{hf}$ yields then a small distribution of the 
magnitude of $H_{eff}$. It entirely accounts for the broadening of the lines with respect
to the zero-field linewidths, which is the most visible at 7\,T. 

\noindent As seen in Fig.\ref{moss4.2}, the model of a random orientation of the moments
satisfactorily reproduces the shape of the M\"ossbauer spectra, except for a small
misfit of the shape of the intermediate pair of lines. This implies that, up to a field
of 7\,T, the magnetic moments of the Fe$^{3+}$ ions remain close to the 
antiferromagnetic axis in our ferritin samples at 4.2\,K. The same qualitative conclusion
can be drawn from the 14\,T spectrum in horse spleen ferritin of Ref.\onlinecite{hunt}.
In the following, we will label this moment configuration as the ``random magnetic
orientation'' configuration.

\subsection{Spectra in the superparamagnetic regime (T=90\,K)}
\begin{figure}
\epsfxsize=7 cm
\centerline{\epsfbox{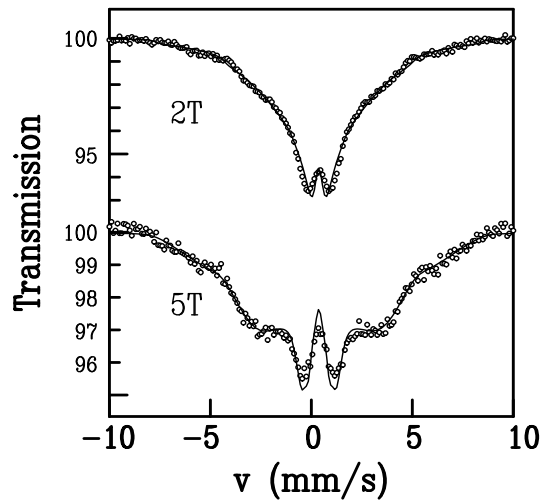}}
\vspace{0.5 cm}
\caption{\small\sl $^{57}$Fe M\"ossbauer absorption spectra in the superparamagnetic 
regime 
(T=90\,K), in the artificial ferritin sample with a mean Fe loading of 410 atoms per core, 
with magnetic
fields of 2\,T and 5\,T applied perpendicular to the $\gamma$-rays propagation direction.
The solid lines correspond to fits using the dynamic ``random magnetic orientation'' model
(see text).}
\label{moss90}
\end{figure}

In the superparamagnetic regime and in zero applied field, the fast 
fluctuations of the hyperfine field smear out the magnetic hyperfine structure.
At 90\,K, the zero field spectrum is a quadrupolar doublet identical with the main
component of the spectrum at 35\,K shown at the bottom of Fig.\ref{mosschmpnul}. 
If a field is applied, the shape of the spectra, shown in Fig.\ref{moss90},
changes, and at 5\,T a six line hyperfine field pattern is
distinguishable, but with large broadenings. 
In order to interpret the spectra, we will
assume that a dynamic ``random magnetic orientation'' model holds at 90\,K, i.e. that the
moments fluctuate along or close to the antiferromagnetic axis. Our analysis follows here
that of Ref.\onlinecite{pier}. For a given orientation $\theta_b$ of the antiferromagnetic 
axis and a given value of the uncompensated moment, the energies of the two particle states 
with ``up'' and ``down'' orientations.of the
uncompensated moment are separated by a Zeeman splitting:
\begin{eqnarray}
\Delta E_z=2\mu_{nc}(V) H \cos\theta_b.
\label{delta}
\end{eqnarray}
At a given Fe site, the hyperfine fields associated with 
each energy level are opposite, resulting in a thermally averaged hyperfine field:
\begin{eqnarray}
H_{hf}(T,V)=H_{hf}^0(T)\tanh[{\mu_{nc}(V) H \cos\theta_b\over k_BT}],
\label{Hhf}
\end{eqnarray}
where $H_{hf}^0(T)$ is the local hyperfine field. The values of $H_{hf}(T,V)$ for a given
particle set are therefore distributed due to both the distribution of $\mu_{nc}(V)$ values 
and to the random distribution in $\theta_b$ values. 

\noindent The fits of the in-field spectra with this dynamic ``random
magnetic orientation'' model, like those of the FC-ZFC curves, are rather insensitive to both
the assumption made about the distribution of uncompensated moment values and the value of
the exponent $p$. For the fits shown as solid lines in Fig.\ref{moss90}, we used the
same distribution of uncompensated moments as for the fits of the FC-ZFC curves, i.e. a 
truncated gaussian shape with a mean value: $\langle \mu_{nc} \rangle = m_0 u \sqrt{N}$. 
We find that the parameter $u$ is 0.4-0.5 for the lowest fields, and that it decreases as 
the field increases. The good reproduction of the spectral shape, especially if one considers
that $u$ is the only adjustable parameter, indicates that it is realistic to consider that 
the individual moments, and hence the sublattice magnetizations, fluctuate along or close to 
the antiferromagnetic axis at 90\,K. The fact that the mean uncompensated moment found here is 
about twice smaller than that derived from the FC-ZFC curves probably originates from the roughness
of the assumption of a gaussian distribution of $\mu_{nc}$ values.
The decrease of $u$ as the field increases, also observed in Ref.\onlinecite{pier}, is probably
due to the fact that the ``random magnetic orientation'' model starts to break down 
at the higher fields, as will be discussed in the next section. The values obtained for 
$u$ at the lowest fields ($\sim$0.5) are similar to those derived from the
magnetization measurements described in section \ref{aimantation}.

\section{Interplay of the exchange, Zeeman and anisotropy energies in uncompensated 
antiferromagnets}
\label{exchange}

Before proceeding with the experimental results of the magnetization measurements and their
interpretation, we will examine the effect of a magnetic field on an uncompensated 
antiferromagnetic particle, and determine the field range for which the ``random magnetic 
orientation'' configuration is expected to be a correct approximation. 
For a fully compensated antiferromagnet, the relevant threshold field is the so-called 
spin-flop field which writes: $H_{sf} \simeq \sqrt{2 H_{\rm A} H_{\rm E}}$, 
where $H_{\rm E}$ is the exchange field and $H_{\rm A}={K\over M_0}$ the anisotropy field,
$M_0$ being the magnetization of one sublattice \cite{neel4}. When the applied field 
{\bf H} is aligned 
along the antiferromagnetic axis, the sublattice magnetic moments rotate to a direction 
perpendicular to it for $H=H_{sf}$. For a particular orientation $\theta_b$ of the 
antiferromagnetic
axis with respect to {\bf H}, the $T$=0 equilibrium direction $\theta$ of the sublattice 
moments,
in the limit $H/H_{\rm E} \ll 1$ and $H_{\rm A}/H_{\rm E} \ll 1$, is given by \cite{beckmann}:
\begin{eqnarray}
cos^2\theta={1\over 2} + {cos^2\theta_b-{1+x^2\over2}\over 
\sqrt{(1+x^2)^2-4x^2cos^2\theta_b}},
\label{costeta}
\end{eqnarray}
where $x={H\over H_{sf}}$.
For all initial orientations, a crossover to a ``quasi spin-flop'' configuration, where
the equilibrium orientations are close to 90$^\circ$,
occurs for $H > H_{\rm sf}$ and, for lower fields, the equilibrium position $\theta$
remains close to the initial orientation $\theta_b$.\par
In order to investigate the effect of the uncompensated moment, we will follow a model
developed by M{\scriptsize$\emptyset$}rup \cite{Morup}. For a particle with volume $V$ and
``degree of uncompensation'' $\alpha = {\mu_{nc} \over {M_0 V}}$, the energy of an 
antiferromagnetic particle is given by:
\begin{eqnarray}
E(\theta, \theta_b) = &M&_0 H_{\rm E} [-(1+\alpha) + {1 \over 2} \beta^2 \sin^2\theta + \cr
&\alpha& \beta \cos\theta + \kappa (1+\alpha) \cos^2(\theta - \theta_b) ],
\label{enernc}
\end{eqnarray}
where $\beta=H/H_{\rm E}$ and $\kappa=H_{\rm A}/H_{\rm E}$. This expression is valid for 
$\beta \ll 1$, $\kappa \ll 1$ and when the ``canting angle'' 2$\epsilon$ of the two 
sublattices has its equilibrium (small) value: 2$\epsilon=\beta \sin\theta$. 
M{\scriptsize$\emptyset$}rup has shown that the spin-flop
field in the presence of an uncompensated moment is enhanced with respect
to $H_{sf}$ \cite{Morup}: 
\begin{eqnarray}
H_{sf}^{nc} = {1 \over 2} \alpha H_{\rm E} + \sqrt{{1 \over 4}
\alpha^2 H_{\rm E}^2 + H_{sf}^2}. 
\label{hsfnc}
\end{eqnarray}
In our ferritin samples, using for instance the values for the sample with 410 Fe atoms
per core: $K \simeq 5 \times 10^5$\,ergs/cm$^3$ and the atomic density:
$N \simeq 1.2 \times 10^{22}$\,Fe$^{3+}$/cm$^3$, the anisotropy field is: $H_{\rm A} 
\simeq$0.18\,T. The exchange field is more difficult to estimate because 
various values for $T_{\rm N}$ have been given in the literature. We will take here
our estimation of the N\'eel temperature, which we show is close to 500\,K for all
particle sizes (see section \ref{aimantation}).
Using the simple molecular field result: $k_BT_{\rm N}={1\over3}g(S+1)H_{\rm E}\mu_B$, 
one obtains an exchange field $H_{\rm E} \simeq 320$\,T. 
As to the values of the degree of uncompensation $\alpha$, they
are probably distributed within a given particle set; using as a reasonable estimate of the
mean uncompensated moment: $\langle \mu_{nc} \rangle$= 0.5 $m_0 \sqrt{N}$, we find that the
mean value of $\alpha$ ranges from 2\% for the biggest particles to 5\% for the smallest ones.

\begin{figure}
\epsfxsize=7 cm
\centerline{\epsfbox{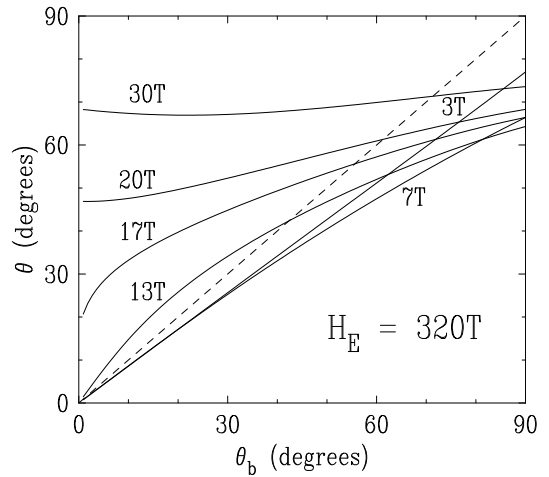}}
\vspace{0.5 cm}
\caption{\small\sl Variation of the equilibrium orientation $\theta$ of the sublattice
magnetizations in an uncompensated antiferromagnet as a function of the orientation 
$\theta_b$ of the
antiferromagnetic axis with respect to the applied field, for H$_{\rm E}$=320\,T. 
The ratio of the uncompensated moment to the sublattice magnetic moment is: $\alpha$=0.03.
The curves (solid 
lines) are calculated for different values of the applied field H ranging from 3\,T
to 30\,T. The dashed line is the curve $\theta=\theta_b$.}
\label{tetahnc}
\end{figure}

\noindent In Fig.\ref{tetahnc}, we show the curves giving the $T=0$ equilibrium 
orientations $\theta$ 
as a function of the initial orientation $\theta_b$, for the representative value 
$\alpha=0.03$, and for different values of the applied field $H$. For this $\alpha$
value, the spin-flop field is: $H_{sf}^{nc}$=16.6\,T. For fields below 7\,T, 
Fig.\ref{tetahnc} shows that
the equilibrium orientation $\theta$ does not strongly depart from $\theta_b$ for all
initial orientations. Thus the $T=0$ ``random magnetic orientation'' model is seen to be 
a good approximation for fields lower than the maximum field of our experiments (7\,T), thus
justifying the use of this approximation to account for the $T$=4.2\,K in-field
M\"ossbauer spectra. This model starts to break down for a field around 7\,T and, above the 
spin-flop field, the 
equilibrium $\theta$ value is seen to be quasi independent of $\theta_b$ and growing
towards 90$^\circ$ as the field is further increased. \par

\begin{figure}
\epsfxsize=7 cm
\centerline{\epsfbox{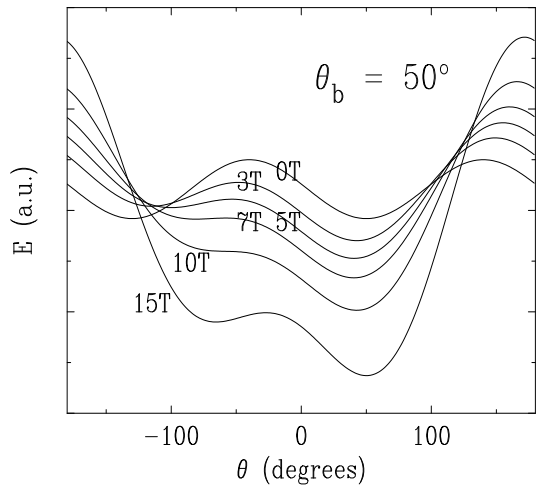}}
\vspace{0.5 cm}
\caption{\small\sl Energy profile for an uncompensated antiferromagnet, for $\theta_b$
=50$^\circ$, a degree of uncompensation $\alpha$=0.03 and an exchange field $H_{\rm E}$=
320\,T, for different values of the applied field. The other parameters 
correspond to the particles with 410 Fe atoms per core.}
\label{enerthnc}
\end{figure}

For a fully compensated antiferromagnetic particle, the energy profiles present two
equally deep potential wells, separated by $\pi$, corresponding to the invariance of the
system by inversion of the two sublattice moments. When the particle possesses an
uncompensated moment, this symmetry is broken and the energy profiles become asymmetrical,
as shown in Fig.\ref{enerthnc} which represents $E(\theta, \theta_b=50^\circ)$ from 
Eqn.\ref{enernc} for different applied fields up to 15\,T, and for $\alpha$=0.03, 
$H_{\rm E}$=320\,T and $H_{\rm A}$=0.18\,T. The two potential wells are still
clearly present for fields below 5\,T, the energy difference between the lowest points
in each well being essentially the Zeeman splitting given by Eqn.\ref{delta}. At higher
fields (7 - 10\,T), the less shallow well is somehow smeared out and it appears anew
when the field is further increased.

\noindent The field range where only one potential well is present is such that 
the Zeeman energy
associated with the uncompensated moment $\mu_{nc} H$ is larger than the anisotropy
energy $K V$, but smaller than the Zeeman energy associated with the canting of the
two antiferromagnetic sublattices, $\chi_\perp H^2$. When $\alpha$ is not too small, this 
yields the approximate boundaries: ${{H_{\rm A}} \over \alpha} < H < \alpha H_{\rm E}$, i.e.
this region lies between 6 and 9.6\,T for $\alpha$=0.03.
At low fields ($H < {{H_{\rm A}} \over \alpha}$),
the positions of the two wells remain close to $\theta_b$ for the shallowest (see 
Fig.\ref{tetahnc}) and $\pi - \theta_b$ for the other. Therefore, in this field region and in 
the superparamagnetic 
regime, the dynamic ``random magnetic orientation'' model, where the uncompensated 
moment is taken to fluctuate along the antiferromagnetic axis, is a fairly good approximation.
At higher fields, the positions and shape of the potentiel wells start to be strongly
modified, which probably accounts for the difficulty of coherently reproducing the M\"ossbauer
spectra at 90\,K using this model.

\section{Isothermal magnetization experiments}\label{aimantation}    

\begin{figure}
\epsfxsize=7 cm
\centerline{\epsfbox{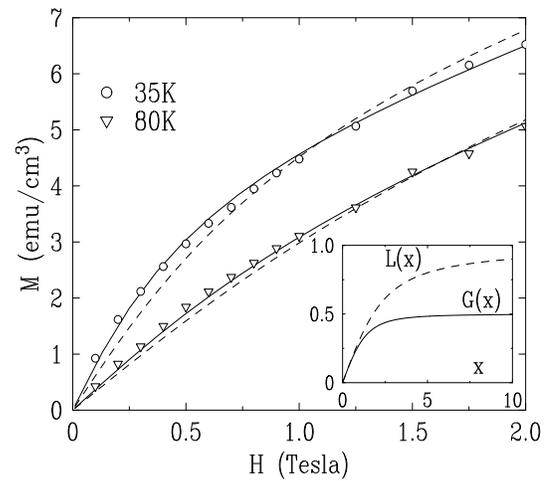}}
\vspace{0.5 cm}
\caption{\small\sl Low field part of the isothermal magnetization at 35\,K and 80\,K in 
the artificial ferritin sample with a mean Fe loading of 982 atoms per core. 
The solid lines are fits 
with the ``random magnetic orientation'' model (see text), and the dashed lines represent 
the fits with the Langevin model. The inset shows the G(x) function (see
text) and the Langevin function L(x).} 
\label{comp}
\end{figure}

\begin{figure}
\epsfxsize=7 cm
\centerline{\epsfbox{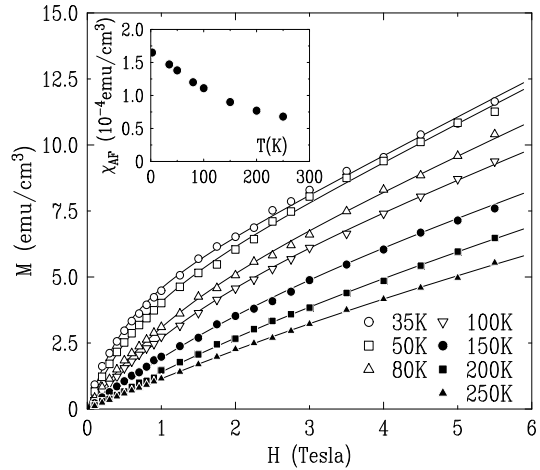}}
\vspace{0.5 cm}
\caption{\small\sl Magnetization curves in the artificial ferritin sample
with a mean Fe loading of 982 atoms per core. The solid lines are obtained by
using the ``random magnetic orientation'' model (see text, Eqn.(\ref{mt}))
The inset shows the thermal variation of $\chi_{\rm AF}$ in the same sample.}
\label{aimh}
\end{figure}

Isothermal magnetization measurements have been performed in the superparamagnetic
regime, between 35\,K and 250\,K, for fields up to 5.5\,T. Representative curves are shown
in Figs.\ref{comp} and \ref{aimh}, for the particles with 982 Fe atoms per core. For 
a given temperature, the magnetization increases with the field, without saturating, and, 
for a given field, it decreases with increasing temperature, as observed in 
Refs.\cite{kilcoyne,berk,brooks,harris}. 

\noindent For these
antiferromagnetic particles, the magnetization can be thought to arise from two 
contributions, as explained in section \ref{superpara}: a linear term $\chi_{\rm AF}H$,
where $\chi_{\rm AF}$ is the powder antiferromagnetic susceptibility and accounts for
the weak canting of the two sublattices, and another term due to the uncompensated
moments. In the previous magnetization studies of ferritin quoted above,
this latter contribution has been interpreted using the Langevin model. However, for such
low fields, the discussion presented in section \ref{exchange} shows that the uncompensated
moments must be considered as fluctuating essentially along the antiferromagnetic axis,
which is confirmed by the analysis of the in-field M\"ossbauer spectra at 90\,K.
This is not compatible with the 
Langevin model, which assumes the moments are free to rotate. In this respect, weakly
uncompensated antiferromagnetic particles strongly differ from ferro- or ferrimagnetic
particles. In the latter, the anisotropy energy is easily overcome by the Zeeman energy and
the Langevin model is a good approximation, although deviations from it have been observed
at low temperature and interpreted using a full Boltzmann 
calculation of the magnetization (see for instance Ref.\onlinecite{hanson}). 
In randomly oriented antiferromagnetic particles, a calculation of the total magnetization
along the same lines is much more difficult because it involves four variables, i.e. two
positional angles for the each of the two sublattices. Therefore, as a low field approximation,
we interpret the contribution of the uncompensated moments in the superparamagnetic regime 
using the dynamic ``random magnetic orientations'' model, assuming that they fluctuate
between two antiparallel directions along the antiferromagnetic axes.
The uncompensated moment, for a given particle volume and a given orientation, is then 
described by a hyperbolic
tangent function similar to the expression for the hyperfine field (Eqn.\ref{Hhf}), and
the measured moment is obtained by integration over the random directions of the 
antiferromagnetic axis:
\begin{eqnarray}
m_{nc}(T,V)=\mu_{nc}(T,V)G[{{\mu_{nc}(T,V) H}\over {k_BT}}],
\label{mnc}
\end{eqnarray}
with the function $G(x)$ defined by:
\begin{eqnarray}
G(x)={1\over 2}\int_0^\pi d\theta \sin\theta \cos\theta \tanh(x \cos\theta).
\label{gx}
\end{eqnarray}
The $G(x)$ function is plotted in the inset of Fig.\ref{comp} together with the Langevin
function $L(x)$; $G(x)$ saturates at the value 1/2, whereas the Langevin law saturates 
at unity. The isothermal magnetization curves have been fitted with this model of dynamic
``random magnetic orientation'', according to the law: 
\begin{eqnarray}
&M&(T,H)=\chi_{\rm AF}(T)H \cr
&+&{1 \over <V>} \int_{V_{min}} ^{V_{max}} \mu_{nc}(T,V) f(V) 
G[{{\mu_{nc}(T,V) H}\over {k_BT}}]dV,
\label{mt}
\end{eqnarray}
where $f(V)$ is the lognormal volume distribution function. We chose to describe the
uncompensated moments by their mean value: $\mu_{nc}(T,V) = m_0 u(T) \sqrt{N}$, where $u(T)$
accounts for the thermal variation of a Fe$^{3+}$ moment.  
In Fig.\ref{comp} are represented the low field parts of the isothermal magnetization
curves in the particles with 982 Fe atoms per core, at 35\,K and 80\,K, together with
the fits of the data using the $G(x)$ function (solid line) and the Langevin function
(dashed line) with the same volume distribution. It is clear that the fit using the
``random magnetic orientation'' model gives a better account of the $M(H)$ curvature
at low fields than does the Langevin fit, the improvement being more pronounced 
at low temperature. 
The complete field and temperature variation of the magnetization in the particles 
with 982 Fe atoms per core
is shown in Fig.\ref{aimh}, the solid lines representing the fit using the 
``random magnetic orientation'' model (Eqn.\ref{mt}). 

\noindent These fits are very satisfactory in the whole field range, and the values for $u$ are
about 0.6 (close to those obtained in the M\"ossbauer spectra at 90\,K at the lowest 
fields) and slowly decrease as temperature increases. We emphasize that the description
of the thermal and field variation of the uncompensated magnetization in the 
superparamagnetic regime using the $G(x)$
function lies on stronger physical grounds than that using a Langevin function.
The upper bound of the field range where this
approach is valid, as discussed in section \ref{exchange}, is approximately $H_{\rm A}
/ \alpha$, i.e. it is close to the maximum field (5--6\,T) used so far in magnetization
studies of ferritin. Above this field, the uncompensated magnetization starts to decrease
and vanishes when the field reaches the spin-flop threshold, as the uncompensated moments
fluctuate along a direction close to perpendicular to the applied field. The $G(x)$
function has been evoked by N\'eel \cite{neel0,neel4} in his studies of antiferromagnetic 
particles: he applied it to calculate the thermoremanent magnetization at the blocking 
temperature and stated that it was the correct description of the uncompensated 
magnetization in the limit of strong anisotropy. This fact has also been noted by Bean
\cite{bean} for the magnetization of ferromagnetic fine powders. We showed here that, in
weakly uncompensated antiferromagnetic particles (i.e. with a degree of uncompensation of 
a few percent), the $G(x)$ function yields a correct description of the magnetization
even for finite and rather small values of the anisotropy energy, like those found in
ferric oxides, for fields up to 5--6\,T. This is finally a consequence of the very large
value of the exchange field, which yields a large spin-flop field.\par

The thermal variation of the mean uncompensated moment derived from our fits of the $M(H)$ 
curves presents a small but significant decrease as temperature increases 
from 35\,K to 250\,K, for all particle sizes. Using an antiferromagnetic magnon law: 
$\mu_{nc}(T)=\mu_{nc}(0)(1-\alpha T^2)$, the experimental $\mu_{nc}(T)$ curves can be
extrapolated to zero in order to yield an estimation of the N\'eel temperature.
We find that $T_N$ is close to 500\,K, and that it is practically the same for
all particle sets. This absence of dependence of $T_N$ on the
mean particle size is probably due to
the fact that the distance between ``active'' reticular planes, which determines
the strength of the superexchange interaction, is independent of the particle size.
The $T_{\rm N}$ value we obtain is roughly two times larger than early estimations
\cite{bell,bauminger}, but is in good agreement with more recent determinations
\cite{pier,berk,brooks} in natural ferritin.\par

The antiferromagnetic susceptibility $\chi_{\rm AF}(T)$, shown in the inset of Fig.\ref{aimh}
for the particles with 982 Fe atoms per core, decreases with increasing temperature, whereas
the bulk powder antiferromagnetic susceptibility is expected to increase on heating.
A mechanism for the thermal decrease of $\chi_{\rm AF}$ has been proposed by N\'eel \cite{neel4},
by considering the reduced exchange field of the superficial Fe layers, but it is 
difficult to say whether it is responsible for the thermal decrease of $\chi _{\rm AF}$
we observe in the ferritin particles.
The $T$=0 value of $\chi_{\rm AF}$, derived from the magnetization curves at 2.5\,K (not shown 
here), amounts to a few 10$^{-4}$\,emu/cm$^3$ for all particle sets.
In the simple molecular field model, the expression for the powder $T$=0 antiferromagnetic 
susceptibility is: $\chi_{\rm AF}(T=0)={2 \over 3} \chi_\perp ={2\over 3}{M_0\over H_{\rm E}}$.
Using the estimated exchange field value $H_{\rm E}$=320\,T, we find that, for each particle
set, the calculated $\chi_{\rm AF}(T=0)$ is smaller than the experimental value
by a factor of about 3. 
N\'eel showed that the susceptibility of antiferromagnetic particles having an even number
of ``active'' reticular planes (i.e. no net magnetic moment) can be enhanced by a factor 
2 or more with respect to the bulk 
susceptibility, and that this is a surface effect visible in very small particles
\cite{neel5}. He called this behaviour ``superantiferromagnetism''. We think that this
mechanism is effective for such particles even in the presence of a small uncompensated 
moment, and that it is responsible for the observed large value of the antiferromagnetic
susceptibility. The superantiferromagnetic behaviour is confirmed by high field (30\,T) 
magnetisation measurements we
performed at 2.5\,K in natural ferritin, which will be the subject of a future
publication \cite{gilles}.\par

At this stage, we would like to comment on the observability of effects linked with surface
atoms in antiferromagnetic particles. Due to the small size of our particles (4--6\,nm),
an important fraction of Fe atoms (around 40\%) lie at the surface, and their properties
(anisotropy, exchange field) are likely to be somehow different from those of the core 
atoms. For instance, in ferro- or ferrimagnetic systems, the observed lack of full alignment
of the magnetic moments with the applied field is characteristic of nanoparticles and
is generally attributed to the different behaviour of surface atoms \cite{kodama,morrish}.
In antiferromagnetic particles in moderate magnetic fields, the situation is different
because the dominant configuration is an almost random orientation of the moments
with respect to the field. Then, effects linked with surface moments are much more
difficult to evidence experimentally and to distinguish from those due to core moments
(see however Ref.\onlinecite{makh}). So, although the properties of surface atoms
may be different from those of the core atoms in our artificial ferritin particles,
we could not detect their effects in our experimental data, except for the 
superantiferromagnetic behaviour mentioned above.

\section{Conclusion}\label{concl}

We performed $^{57}$Fe M\"ossbauer absorption spectroscopy and magnetization
measurements in antiferromagnetic artificial ferritin particles, with mean Fe loadings 
ranging from 400 to 2500 atoms, in the temperature range 2.5\,K-250\,K and with magnetic fields
up to 7\,T. In zero or very low field, the dynamics of the sublattice magnetization of the 
ferritin particles obeys classical
superparamagnetic relaxation in the thermal activation regime. The value of the anisotropy 
energy per unit volume $K$ could be determined from the FC-ZFC susceptibility curves
and from the thermal variation of the M\"ossbauer superparamagnetic fraction. The
values obtained by both techniques are in good agreement and are in the range 
$3-6 \times10^5$\,ergs/cm$^3$, typical for ferric oxides or hydroxides, with a tendency
for $K$ to increase as the particle size decreases. 
By comparing the data obtained by M\"ossbauer spectroscopy and by magnetometry in moderate
magnetic fields up to 7\,T, we propose a new interpretation of the field and temperature 
behaviour
of the uncompensated magnetization in the superparamagnetic regime. We show that it
is better described by a ``random magnetic orientation'' model, where the uncompensated
moments fluctuate along the antiferromagnetic axis, than by the usually assumed Langevin 
law. This behaviour is described by the universal function: 
\begin{eqnarray}
G(x)={1\over 2}\int_0^\pi d\theta \sin\theta \cos\theta \tanh(x \cos\theta).
\label{gx1}
\end{eqnarray}
To our knowledge, this law has never been applied to the interpretation of data in
antiferromagnetic particles, although N\'eel had noticed that it should be the correct
description of the thermal and field variation of the uncompensated magnetization
in the limit of large anisotropy \cite{neel4}.
We think this non-Langevin behaviour actually holds for usual values of the anisotropy
energy, and it should be valid not only in ferritin, but
also in other weakly uncompensated antiferromagnetic particles, at least
in moderate magnetic fields. As the field is increased towards the spin-flop field 
(10--20\,T) and above, the ``random magnetic orientation'' picture breaks down, as the 
magnetic moments progressively reorient perpendicular to the applied field and the 
superparamagnetic uncompensated magnetization vanishes.\par

We could also estimate the N\'eel temperature in our artificial ferritin samples by
following the decrease of the mean uncompensated moment in the temperature range
35--250\,K; we find that $T_{\rm N}$ is essentially independent of the mean particle
size, and that it is close to 500\,K, in agreement with other recent estimations.\par   

\bigskip

The authors are grateful to Dr E. Vincent and Dr G. LeBras, from the Service de Physique
de l'Etat Condens\'e (CEA Saclay), for their help with the SQUID measurements.

\end{multicols}

\end{document}